\documentclass[10pt,a4paper,prd,tightenlines,nofootinbib,showpacs,showkeys,twocolomn, onecolomn, superscriptaddress, notitlepage,reprint]{revtex4-1}
\usepackage{bm}

\usepackage[pdftex]{graphics}
\usepackage{rotating}
\usepackage{epsfig}
\usepackage[usenames]{color}
\usepackage{epstopdf}
\usepackage{mathtext}
\usepackage{amsmath}

\begin{document}
\title{New noise-induced statistically steady state in the spatially extended competition model}

\author{\firstname{S.\,E.} \surname{Kurushina}}
\affiliation{Physics Department, Samara State Aerospace University named after S.P. Korolyov, Moskovskoye Shosse 34, 443086,
Samara, Russian Federation}
\author{ \firstname{V.\,V.} \surname{Maximov}}
\affiliation{Physics Department, Samara State Aerospace University named after S.P. Korolyov, Moskovskoye Shosse 34, 443086,
Samara, Russian Federation}
\author{ \firstname{E.\,A.} \surname{Shapovalova}}
\affiliation{Physics Department, Samara State Aerospace University named after S.P. Korolyov, Moskovskoye Shosse 34, 443086,
Samara, Russian Federation}
\author{\firstname{Yu.\,M.} \surname{Romanovskii}}
\affiliation{Physics Department, Lomonosov Moscow State University, GSP-1, Leninskie Gory, 119991,
Moscow, Russian Federation}
\author{ \firstname{I.\,P.} \surname{Zavershinskii}}
\affiliation{Physics Department, Samara State Aerospace University named after S.P. Korolyov, Moskovskoye Shosse 34, 443086,
Samara, Russian Federation}
\author{\firstname{D.\,S.} \surname{Garipov}}
\affiliation{Mathematics Department, Samara State Transport University, First Bezimyannii Pereulok 18, 443066 Samara, Russian Federation}

\begin{abstract}
The influence of an external random field on the competition process in a nonlinear open spatially extended system is analyzed numerically. A three-component model is chosen as the competition model in which a "weak" species can move in space and the rate of the resource density growth fluctuates in space and in time [A.\,S. Mikhailov and I.\,V. Uporov, Usp. Sov. Phys. Usp. \textbf{27}, 695 (1984)]. It is demonstrated that in addition to the noise-induced statistically steady state found by the authors [Sov. Phys. Usp. \textbf{27}, 695 (1984)], in which both species can coexist, there exists another noise-induced statistically steady state, in which the "weak" species displaces the "strong" species, i.e. the "strong" species at an average asymptotically disappears.
\end{abstract}

\pacs{05.40.-a, 02.70.Bf, 87.18.Tt}

\maketitle

\twocolumngrid{

\section{Introduction}

Many physical, chemical, economic, social, and ecological problems lead to the investigation of competition processes between the interacting components of the system.

Competition processes play an important role in the course of evolution of nonlinear open spatially extended systems. For example, the spatial and spatio-temporal pattern formation can be considered as a process of interaction and competition between unstable modes of the system, as a result of which one or a small number of such modes subjugate all others \cite{Haken1}.

Noises have a no less important and quite nontrivial influence on the evolution of nonlinear open spatially extended systems. In addition to the effects investigated in \cite{LiuJin2,GosMar3,RiazDut4,Lindn5,KawSail6,BucIba7,Zhou8, SanzZhab9, SegShap10, ZimTor11, SanSan12, IbaGar13, WanJun14, NeiPei15, ZaikSchim16, GenSan17, GamHan18, SantCol19, ElsSel20, BroPar21, ParBro22, GarSan23, MarGam24, BroPar25, GarHer26,Deis27,Kur28,Kur29},  they can lead to the appearance of new noise-induced statistically steady states \cite{Mikh30}.

The purpose of the present paper is to study numerically the influence of external noise on the competition process in a nonlinear open spatially extended system. We will not consider the general case here, but confine ourselves to the study of a simple, but biologically important model. The outline of the rest of the paper is as follows. The model under study is presented in Sec. II. The results obtained in Ref.\cite{Mikh30} are briefly discussed. The numerical method used for simulation is presented in Sec. III. The possibility of its application to the problem under consideration is established. The results of simulation of competition processes for different values of some parameters of the problem are presented in Sec. IV. Three types of solutions found are described. One type of solution is new and it corresponds to the situation, when an initially "strong" species surrenders in competitive fighting and disappears. Finally, some conclusions are reported in Sec. V.

\section{The model}

In the paper \cite{Mikh30} a model of the Volterra type describing the interaction of two biological species relying on the same resource was introduced. It is assumed that individuals of one species are able to move in space, which is modeled by the diffusion term in the appropriate equation and the rate of the resource density growth changes randomly in space and in time. The model equations are as follows:
\[
{\frac{\partial s}{\partial t}} = (B r-A)s,
\]
\begin{equation}
\label{eq1}
{\frac{\partial w}{\partial t}} = (b r-a)w + D \nabla ^2 w,
\end{equation}
\[
{\frac{\partial r}{\partial t}} = Q + f(\textbf{r},t) - Gr - Cs - cw,
\]
where $s, w$ are the population densities of "strong" and "weak" species respectively, $r$ is the resource density; $A,a (B, b)$ are the coefficients of natural change of population; $Q$ is the rate of resource growth; $C, c$ are the coefficients of its consumption; $G$ is the coefficient of the natural decline of resource. The term $D{\nabla}^2w$ takes into account the mobility of individuals of the "weak" species. The random field $f(\mathbf{r},t)$ with zero mean defines spatial and temporal fluctuations of resource density growth. All the coefficients in Eq.~(\ref{eq1}) are positive. It is additionally assumed that the conditions $A/B < a/b, Q > GA/B$ are fulfilled.
\par As noted in \cite{Horst31}, fluctuations in the environment represent the summarized effect of many weakly coupled factors. Therefore, according to the central limit theorem fluctuations of the external source have a Gaussian distribution. The ergodic Markovian and Gaussian properties of the fluctuating environment limit the choice of random fields for modeling the fluctuations of the environment by a stationary homogeneous isotropic Gaussian field with the exponential time- and space-correlation function. Therefore, in this paper as in \cite{Kur28}

\begin{equation}
\label{eq2}
\langle f(\textbf{r},t)f(\textbf{r}',t') \rangle = 2G \theta \exp(-k_{f}|\textbf{r}-\textbf{r}'|)\exp(-k_{t}|t-t'|)
\end{equation}

Here $r_f=k_f^{-1}$ determines the characteristic spatial scale of fluctuations, $r_t=k_t^{-1}$ determines the characteristic temporal scale of fluctuations, $\theta$ is their intensity. The correlation time is significantly shorter than all characteristic times of the problem. In the paper \cite{Mikh30} field $f(\mathbf{r},t)$ is a $\delta$-correlated in time.
\par The only stable solution of the local deterministic system (\ref{eq1}), defined as
\[
w_s=0, r_s=A/B, s_s=(Q-Gr_s)/C
\]
corresponds to Gause competitive exclusion principle \cite{Gause32}.

It was shown in \cite{Mikh30} that the situation becomes different from the classical one if the rate of resource density fluctuates in space and in time. Beginning with some critical noise intensity stationary statistical coexistence of two competing species becomes possible. The authors \cite{Mikh30} named this phenomenon "medium populating". "Weak" species population density average with respect to volume and asymptotic over time becomes equal to:
\[
{{\langle w \rangle}_{V}}_{s}=
{\left\{
{\begin{array}{*{20}l}
 {0,\theta < \theta _{c} ;} \hfill \\
 { (b{p}_{1}/R)(1/{\theta}_{c} - 1/\theta),\theta > \theta _{c} } \hfill \\
\end{array}}
\right.}
\]
Here $R=3{\sqrt{2}}b^{3}c/[4G {w}^2_{0}{(D{k}^2_{f})}^{3/2} ],$$\omega_{0}=$$\sqrt{B(Q-GA/B)},$ $\theta_{c}=p_{1}D{k}^2_{f}/b, p_{1}=a/b - A/B.$ The latter value determines the resource deficit for the reproduction of "weak" species individuals in the steady state. Herewith it is also shown in \cite{Mikh30} that without diffusion fluctuations of the resource growth rate do not prevent asymptotic extinction of "weak" species, therefore this kinetic transition is fundamentally associated with the presence of diffusion. Thus, in conditions of fluctuating environment mobility is the factor ensuring the survival of the species.

Analyzing the system (\ref{eq1}) analytically  the authors of paper \cite{Mikh30} used a number of restrictions  significantly narrowing the range of applicability of the obtained results: complex hierarchy of microscopic scales with the dimension of reverse time, restrictions on the noise intensity, the smallness of  deviations of the concentrations $w$ and $s$ from the stationary values, the vicinity to the transition point to the regime with non-zero  "weak" species population density average with respect to volume ${{\langle w \rangle}_{V}}\ne0$ etc. In reference with the above carrying out numerical analysis of the system~(\ref{eq1}) evolution in the absence of the limitations above appears to be interesting.\\

\section{The numerical method}

Let us focus on the description and justification of the possibility of applying the numerical method used for the simulation of the system (\ref{eq1}) evolution in more detail.

We assume that species interaction occurs in a large, but finite area of space. Let us consider a one-dimensional problem. Then the system of equations (\ref{eq1}) with account for normalization can be rewritten as follows:
\[
{\frac{\partial \widetilde{s}}{\partial \tau}} = (\widetilde{r}-1)\widetilde{s},
\]
\begin{equation}
\label{eq3}
{\frac{\partial \widetilde{w}}{\partial \tau}} = \left( \frac{b}{B} \widetilde{r}-\frac{a}{A} \right) \widetilde{w} + \frac{D}{A} \frac{\partial^2 \widetilde{w}}{\partial x^2},
\end{equation}
\[
{\frac{\partial \widetilde{r}}{\partial \tau}} = \frac{B}{A^2} \left[ Q + f(x,\tau) - \frac{AG}{B}\widetilde{r}
- \left(Q-G\frac{A}{B}\right)(\widetilde{s} + \frac{c}{C}\widetilde{w}) \right],
\]
where $\tau=At; \widetilde{s}=s/s_s; \widetilde{r}=r/r_s; \widetilde{w}=w/s_s$. The fixed boundaries are assumed to be impermeable:
\[
{\left.{\frac{\partial \widetilde{s}}{\partial x} }  \right|}_{x=0;L}=0,
{\left.{\frac{\partial \widetilde{w}}{\partial x} }\right|}_{x=0;L}=0,
{\left.{\frac{\partial \widetilde{r}}{\partial x} }\right|}_{x=0;L}=0,
\]
where $L$ is the characteristic size significantly exceeding all characteristic spatial scales of the problem.

The scheme of the numerical method used for integration (\ref{eq3}) is written on the basis of the following. Let us formally write the solution of the system (3) in its equivalent integral form:
\[
\widetilde{s}= {\widetilde{s}}_{0} + \int \limits_{t_{0}} ^{t} (\widetilde{r}-1)\widetilde{s} d \tau,
\]

\begin{equation}
\label{eq4}
\widetilde{w}= {\widetilde{w}}_{0} + \int \limits_{t_{0}} ^{t}{ \left[ \left(\frac{b}{B} \widetilde{r}-\frac{a}{A}\right)\widetilde{w} + \frac{D}{A} \frac{\partial^2 \widetilde{w}}{\partial x^2}   \right]} d \tau,
\end{equation}
\[
\widetilde{r}= {\widetilde{r}}_{0} + \frac{B}{A^2} \int \limits_{t_{0}} ^{t}\left[ Q - \frac{AG}{B}\widetilde{r}- \left(Q-G\frac{A}{B}\right)(\widetilde{s} + \frac{c}{C}\widetilde{w}) \right] d \tau
\]
\[
                      \qquad + \frac{B}{A^2} \int \limits_{t_{0}} ^{t} f(x,\tau)d \tau
\]
Real noise as opposed to white one has realizations continuous at almost all points, therefore, due to the smoothing effect of integration the process with the realizations differentiable at almost all points will correspond to the solution (\ref{eq4}) of the system of stochastic differential equations (SDE) (\ref{eq3}). That is why  (\ref{eq3}) can be interpreted as a system of ordinary differential equations for realizations and $ \int\limits_{t_0}^t f(x,\tau) d\tau$ from the third equation ~(\ref{eq4}) can be understood in the sense of the  Riemann integral.

All the aforesaid makes it possible to use the conventional two-layer finite-difference scheme  for SDE (\ref{eq3}) in which $f(x,\tau)$ can be interpreted as part of the nonlinear function in the right-hand side. Then we obtain on a rectangular uniform grid $[0\le x\le L]\times[0\le \tau\le T]$ with the time step $\Delta\tau$ and space step $h$:
\[
\widetilde{s}^{j}_{i}= \Delta \tau (\widetilde{r}^{j-1}_{i} -1)\widetilde{s}^{j-1}_{i} +\widetilde{s}^{j-1}_{i}
\]
\begin{equation}
\label{eq5}
{\begin{array}{l}
\widetilde{w}^{j}_{i-1}- \left( 2+\frac{Ah^2}{D \Delta \tau \sigma}  \right)\widetilde{w}^{j}_{i} + \widetilde{w}^{j}_{i+1}= \\
\qquad -\frac{1-\sigma}{\sigma}\left(\widetilde{w}^{j-1}_{i-1}-2\widetilde{w}^{j-1}_{i} +\widetilde{w}^{j-1}_{i+1}  \right)-\frac{Ah^2}{D \Delta \tau \sigma}\widetilde{w}^{j-1}_{i}\\
\qquad -\frac{Ah^2}{D \Delta \tau \sigma}\left(\frac{b}{B}\widetilde{r}^{j-1}_{i} -\frac{a}{A}  \right)\widetilde{w}^{j-1}_{i}\\
\end{array}}
\end{equation}
\[
\begin{array}{l}
\widetilde{r}^{j}_{i} = \Delta \tau \frac{B}{A^2} \left[ Q -
\frac{AG}{B}\widetilde{r}^{j-1}_{i} - \left( Q-G\frac{A}{B} \right)  \left(\widetilde{s}^{j-1}_{i}+
\frac{c}{C}\widetilde{w}^{j-1}_{i}\right)   \right]\\
 \qquad + \Delta \tau \frac{B}{A^2}{f}^{j-1}_{i} + \widetilde{r}^{j-1}_{i}
 \end{array}
\]
Here $f_i^j$ are realizations of a random Gaussian field with the appropriate correlation function, $\sigma=1/2$ is the weighting factor of the scheme at the spatial derivative from the upper layer.

The realizations of the field were obtained from the following considerations. Let us suppose that time dependence of the field could be considered practically the same at all points in space. Then we can write $f(x,\tau)=u(x)v(\tau)$. For such a field the correlation function takes the form $B(x,\tau)=B_u(x)B_v(\tau)$ that corresponds to the form (\ref{eq2}). Further processes $u_i^j$ and $v^j$ are implemented according to the scheme:
\[
 \begin{array}{l}
 u^{j}_{i}= [\theta_{1}(1-\exp(-2k_{f}|x_{i}-x_{i-1}|))]^{1/2}e_{i}  \\
       \qquad   + u^{j}_{i-1}\exp(-k_{f}|x_{i}-x_{i-1}| ),
           \end{array}
\]

\[
 \begin{array}{l}
 v^{j}= [\theta_{2}(1-\exp(-2k_{t}|t^{j}-t^{j-1}|))]^{1/2}e^{j} \\
          \qquad +v^{j-1}\exp(-k_{t}|t^{j}-t^{j-1}| )
          \end{array}
\]
$\theta_1\theta_2=2G\theta$, $u_0^j$  is a random Gaussian number with zero expectation and variance $\theta_1$, $v^0$ is a random Gaussian number with zero expectation and variance $\theta_2$, $e_i$  and  $e^j$  are  random Gaussian numbers with zero expectations and unit variances.  Then $f_i^j=u_i^jv^j$.
Difference boundary conditions are determined by the expressions

\[   \widetilde{w}^{j}_{-1}= \widetilde{w}^{j}_{1} , \widetilde{w}^{j}_{max i-1}= \widetilde{w}^{j}_{maxi+1}  \]
\begin{equation}
\label{eq6}
 \widetilde{s}^{j}_{-1}= \widetilde{s}^{j}_{1} , \widetilde{s}^{j}_{max i-1}= \widetilde{s}^{j}_{maxi+1}
 \end{equation}
\[   \widetilde{r}^{j}_{-1}= \widetilde{r}^{j}_{1} , \widetilde{r}^{j}_{max i-1}= \widetilde{r}^{j}_{maxi+1}  \]
The system~(\ref{eq5}) with boundary conditions~(\ref{eq6}) is solved by the tridiagonal matrix algorithm or any other method of solving systems of linear algebraic equations.

The second equation of the scheme~(\ref{eq6}) represents six-point difference scheme of the Crank - Nicholson  type. The first and third equations are implementations of the simplest Euler scheme.

During the simulation of system (\ref{eq1}) dynamics steady statistical characteristics are determined for parameter values close to bifurcation. As the system approaches the bifurcation point the phenomenon of critical slowing-down is observed due to which the achievement of a statistically steady state requires large intervals of model time. Therefore in simulating the evolution of (\ref{eq1}) it is necessary to keep to the asymptotic stability condition for the second equation of scheme (\ref{eq5}).

\section{The results of simulation}

By using the scheme~(\ref{eq5}) we studied the change of the system~(\ref{eq3}) dynamics depending on the values of parameters $D, p_1,$ and $\theta$. Volume averaged species population densities
\[{{\langle\widetilde {w}(t) \rangle}_{V}}=\frac{1}{L}\int\limits_0^L \widetilde {w}(x,t) dx,
{{\langle\widetilde {s}(t) \rangle}_{V}}=\frac{1}{L}\int\limits_0^L \widetilde {s}(x,t) dx\]
 and them statistically steady values
\[{{\langle\widetilde {w} \rangle}_{Vs}}=\lim\limits_{t\to\infty}{\langle\widetilde {w}(t) \rangle}_{V},
{{ \langle\widetilde {s}\rangle}_{Vs}}=\lim\limits_{t\to\infty}{\langle\widetilde {s}(t) \rangle}_{V},\]
were chosen as values characterizing this change. In Fig.~\ref{fig1}  the parametric diagrams on the planes of parameters $ D - p_1$ and $D - \theta$ are presented. The regions corresponding to different outcomes of competitive fighting are distinguished. The figure shows that there are three different outcomes for system~(\ref{eq3}) -- the three regimes of behavior, with the region of the U-shaped form on both planes corresponding to one of them.

\begin{figure}[h!]
\includegraphics[width=2.345in,height=2.73in]{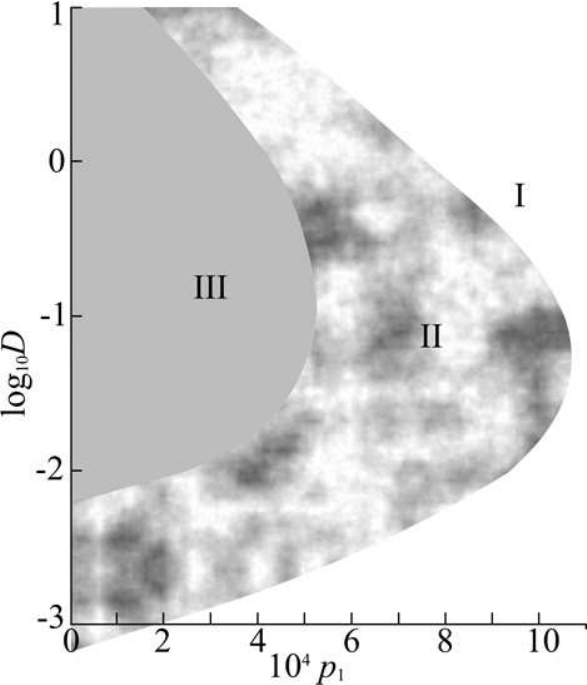}\\
(a)\\
\includegraphics[width=2.438in,height=2.797in]{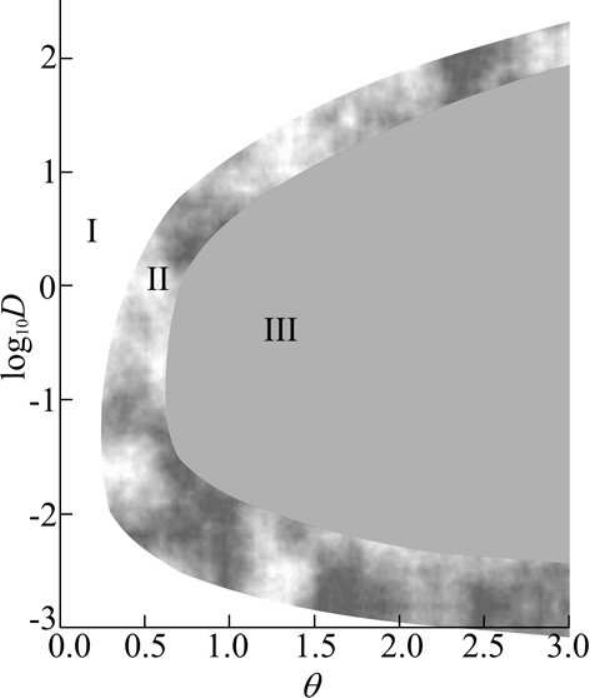}\\
(b)\\
\caption{\label{fig1}Parametric diagrams of the system~(\ref{eq3}) (a) on plane $D - p_1$; (b) on plane $D - \theta$.  I is the region of classical solution (Gause principle); II is the region of the "medium populating" regime; III is the region of the "inversion" regime.}
\end{figure}
For the parameters belonging to region I in the Figs.~\ref{fig1} a,b the outcome determined by the Gause competitive exclusion principle is realized: the population density of the "weak" species  ${{\langle\widetilde {w}\rangle}_{Vs}}$ asymptotic over time and average with respect to volume  tends to zero, i.e. "weak" species disappears.

"Medium populating" by individuals of  "weak" species, predicted in paper \cite{Mikh30} is observed in the parameter region corresponding to the region II in Figs.~\ref{fig1} a, b.  A statistically steady state is established wherein the population densities of species ${{\langle\widetilde {w} \rangle}_{Vs}}$, ${{\langle\widetilde {s} \rangle}_{Vs}}$, average with respect to the volume  and asymptotic over time are different from zero. Typical time dependencies of volume averaged densities ${{\langle\widetilde {w}(t) \rangle}_{V}}$ and ${{\langle\widetilde {s}(t) \rangle}_{V}}$ in region II are presented in Fig.~\ref{fig2}a.

\begin{figure}[h!]
\includegraphics[width=3.11in,height=1.773in]{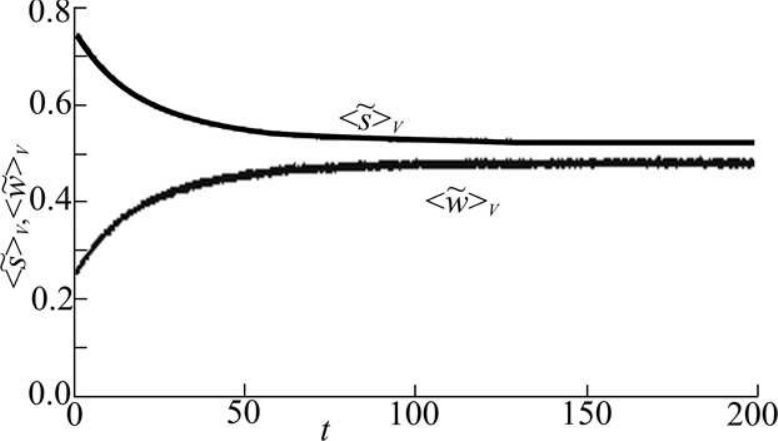}\\
(a)\\
\includegraphics[width=3.105in,height=1.77in]{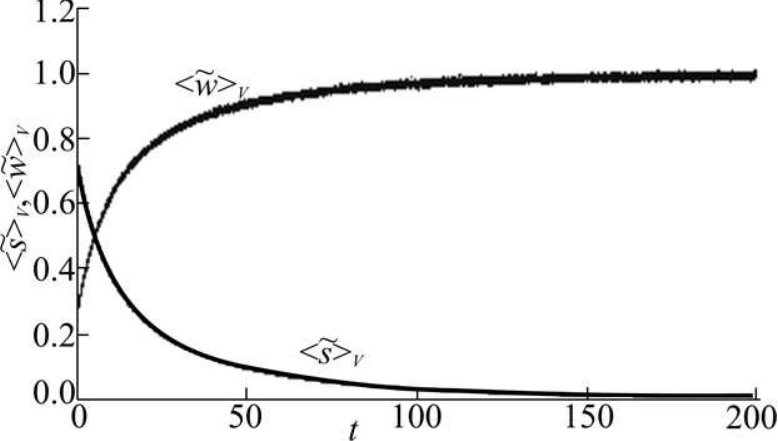}\\
(b)\\
\caption{\label{fig2}Typical time dependences of volume averaged population densities of "weak" species   ${{\langle\widetilde {w}(t) \rangle}_{V}}$ and  "strong" species ${{\langle\widetilde {s}(t) \rangle}_{V}}$ .(a) The "medium populating" regime $D=0.01$. (b) The "inversion" regime $D=0.5$. The other parameters of the model are $A=B=1; a=4.755; b=4.752; C=c=1; Q=9.25; G=3.68; \theta=0.7; k_f=5.4$.}
\end{figure}
Our studies have shown that the model~(\ref{eq3}) has yet another previously unknown regime of behavior wherein  ${{\langle\widetilde {w} \rangle}_{Vs}}\ne0$ and  ${{\langle\widetilde {s}\rangle}_{Vs}}\to0$. That is, the "strong" species on the average asymptotically disappears, conceding to the "weak" species in the competition process (region III in Figs.~\ref{fig1} a,b). In this parameter region "inversion" of properties of the species occurs: a "weak" species becomes a "strong" one. Thus, in conditions of fluctuating environment mobility is a factor, which not only ensures the survival of the species, but is a margin of victory in competitive fighting. Typical dependencies ${{\langle\widetilde {w}(t) \rangle}_{V}}$ and ${{\langle\widetilde {s}(t) \rangle}_{V}}$ in region III are presented in Fig.~\ref{fig2}b. \

The peculiarities of system behavior according to parameters $D, p_1$ are quite understandable. With increasing resource deficit available to "weak" species Gause principle is realized "weak" motionless species disappearing under condition $D\to0$ and $p_1\to0$ with an arbitrarily small resource deficit. Decreasing resource deficit available to "weak" species leads to smoothing out the differences in the dynamics of "weak"  and "strong" species and mobility provides additional competitive advantages for "weak" species. As a result, at first the possibility of coexistence of species emerges and then the "inversion" regime comes into being as $p_1$ decreases. The plots reflecting the results of the competitive process depending on parameter $p_1$ are presented in Fig.~\ref{fig3}a.

\begin{figure}[h!]
\includegraphics[width=2.74in,height=1.955in]{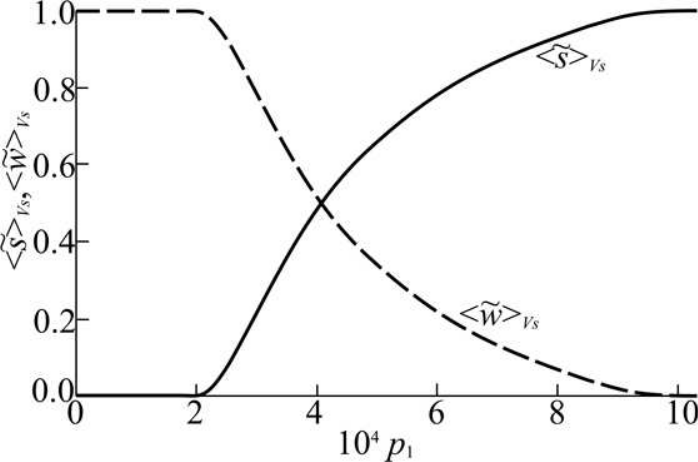}\\
(a)\\
\includegraphics[width=2.793in,height=1.845in]{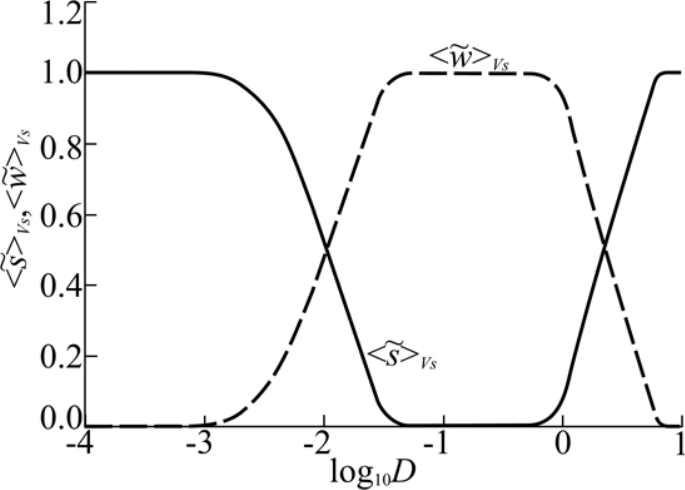}\\
(b)\\
\includegraphics[width=2.747in,height=1.943in]{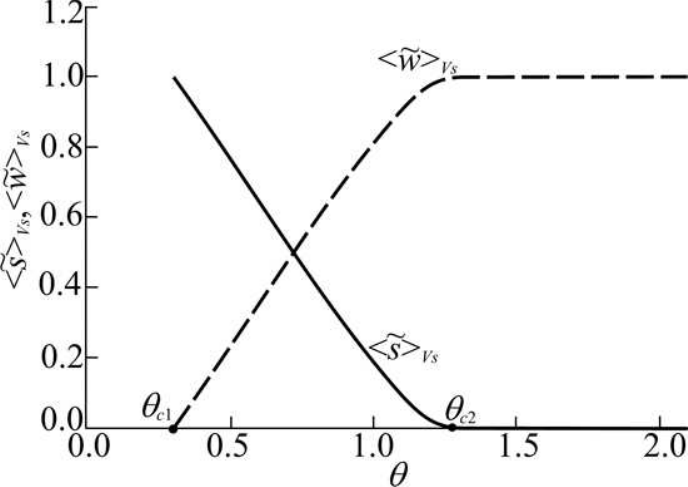}\\
(c)\\
\caption{\label{fig3}Dependencies of  population densities of "weak" species ${{\langle\widetilde {w} \rangle}_{Vs}}$ and "strong" species ${{\langle\widetilde {s} \rangle}_{Vs}}$ asymptotic over time and average with respect to volume (a) on resource deficit $p_1$; (b) on the diffusion coefficient $D$; (c) on noise intensity $\theta$. Other model parameters are the same as in Fig.~\ref{fig2}.}
\end{figure}
Let us remark here that the "inversion" regime is threshold over the parameter $D$ (see Fig.~\ref{fig3}b): under condition $D\to D_с(p_1,\theta)$ additional competitive advantages disappear. Increase of the mobility coefficient $D$ in the limit of large values leads to the fact that the last term in the second equation of system~(\ref{eq3}) becomes prevailing. In the case of high mobility individuals of the  "weak" species pass through the region with resource surplus too fast and cannot use it effectively. In the asymptotics under condition $D\to\infty$ this equation admits of damped solutions of the diffusion type regardless of the values of resource deficit. This, in particular, explains the U-shaped form of region II in Fig.~\ref{fig1}.

The separation of regimes on the plane $D-\theta$ is also easily explained. The smaller the intensity of the fluctuations the smaller the resource available exclusively to the "weak" species and randomly arising in randomly distributed areas of space. The possibility of coexistence of species arises when the intensity fluctuation $\theta$ is higher than the first critical value $\theta_{c1}$. The "weak species" gains an advantage when the intensity fluctuation $\theta$ is higher than the second critical value $\theta_{c2}$ since this type of resource is directly accessible only to it and the possibility of "inversion" arises. The change of densities of the number of "weak" species ${{\langle\widetilde {w} \rangle}_{Vs}}$ and "strong" species ${{\langle\widetilde {s}\rangle}_{Vs}}$ asymptotic over time and average with respect to volume according to value $\theta$ is presented in Fig.~\ref{fig3}c.\\

\section{Conclusion}

In our paper we have studied numerically the influence of external real noise on the competition processes in a nonlinear spatially extended system. Our studies have shown that model~(\ref{eq1}) admits of three different outcomes of competitive fighting: classical disappearance of the "weak" species; noise-induced "medium populating" by individuals of "weak" species predicted in \cite{Mikh30}; as well as a nontrivial outcome -- noise-induced extinction of initially "strong" species. Some parameter region of the problem corresponding to the above competition results  were determined. It is reflected on the parametric diagrams in Fig.~\ref{fig1}.

The studies presented are interesting in that they explain at least some reasons because of which a nontrivial result of competitive fighting occurring in a fluctuating environment is possible. They can also help determine the strategy and tactics of behavior of any competing communities aimed at winning in competitive fighting, including communities of a nonbiological origin.

\section*{Acknowlegment}

The study has been supported by the Ministry of Education and Science of the Russian Federation, Competitiveness Enhancement Program of SSAU for 2013-2020 years and Project No.102 of State assignment to educational and research institutions; Grants No. 13-01-970050 r\_povolzhie\_a of the Russian Foundation for Basic Research.

}
\end{document}